\documentclass[structabstract]{aa}
\usepackage[latin1]{inputenc}
\usepackage{amsmath}
\usepackage{amsfonts}
\usepackage{amssymb}
\usepackage{graphicx} 
\usepackage{subfigure}
\usepackage{natbib}
\usepackage{txfonts}
\usepackage{hyperref}

\begin{document}

 \title{High Frame-rate Imaging Based Photometry\thanks{based on observation with the Danish 1.54m telescope at ESO La Silla Observatory.}}
\subtitle{Photometric Reduction of Data from Electron-multiplying Charge Coupled Devices (EMCCDs)}
\titlerunning{High Frame-rate Imaging Photometry}

   \author{Kennet B.W. Harps\o e\inst{1,2}, Uffe G. J\o rgensen\inst{1,2}, Michael I. Andersen\inst{1} \and Frank Grundahl\inst{3}}
   
   \institute{Niels Bohr Institute, University of Copenhagen,
              Juliane Maries Vej 30, 2100 K\o benhavn \O, Denmark\\
              \email{harpsoe@nbi.ku.dk}
           \and
              Centre for Star and Planet Formation, Natural History Museum of Denmark, University of Copenhagen, \O ster Voldgade 5-7, 1350 K\o benhavn K, Denmark\\
           \and
              Department of Physics and Astronomy, Aarhus University\\ 
              Ny Munkegade 120, 8000 Aahus C, Denmark}

   %\date{Received September 15, 1996; accepted March 16, 1997}

% \abstract{}{}{}{}{} 
% 5 {} token are mandatory
 
  \abstract
  % context heading (optional)
  % {} leave it empty if necessary  
   {The EMCCD is a type of CCD that delivers fast readout times and negligible readout noise, making it an ideal detector for high frame rate applications which improve resolution, like lucky imaging or shift-and-add. This improvement in resolution can potentially improve the photometry of faint stars in extremely crowded fields significantly by alleviating crowding. Alleviating crowding is a prerequisite for observing gravitational microlensing in main sequence stars towards the galactic bulge.  However, the photometric stability of this device has not been assessed. The EMCCD has sources of noise not found in conventional CCDs, and new methods for handling these must be developed.}
  % aims heading (mandatory)
   {We aim to investigate how the normal photometric reduction steps from conventional CCDs should be adjusted to be applicable to EMCCD data. One complication is that a bias frame cannot be obtained conventionally, as the output from an EMCCD is not normally distributed. Also, the readout process generates spurious charges in any CCD, but in EMCCD data, these charges are visible as opposed to the conventional CCD. Furthermore we aim to eliminate the photon waste associated with lucky imaging by combining this method with shift-and-add.}
  % methods heading (mandatory)
   {A simple probabilistic model for the dark output of an EMCCD is developed. Fitting this model with the expectation-maximization algorithm allows us to estimate the bias, readout noise, amplification, and spurious charge rate per pixel and thus correct for these phenomena. To investigate the stability of the photometry, corrected frames of a crowded field are reduced with a PSF fitting photometry package, where a lucky image is used as a reference.}
  % results heading (mandatory)
   {We find that it is possible to develop an algorithm that elegantly reduces EMCCD data and produces stable photometry at the 1\% level in an extremely crowded field.}
  % conclusions heading (optional), leave it empty if necessary 
   {}

   \maketitle

\section{Introduction}

There are a number of exciting areas of astrophysical research that could benefit from accurate, precise, high time- or angular-resolution photometry in crowded fields.  For instance, the search for Earth-mass exoplanets in gravitational microlensing events calls for photometry with a precision of order 1-2\% in the crowded stellar fields of Baade's window \citep{Joergensen2008}.

As a detector for light in the optical and UV parts of the electromagnetic spectrum, the CCD is ubiquitous in astronomy. CCDs have a high quantum efficiency and low dark current when cooled appropriately. Under optimal conditions, the dominant source of noise in the CCD itself is the readout noise. With low-noise CCDs the readout noise can be as low as 2-3 electrons, using very slow readout speeds ($\approx 5\cdot 10^{4}$ pixel per second). With higher readout speeds, above $\approx 10^{6}$ pixels per second, the readout noise increases to beyond ten electrons per readout.

By recording frames at a high frame-rate, one can reduce the impact of atmospheric seeing \citep{Fried78}, using methods such as lucky imaging and shift-and-add. But even with the very lowest readout noise achieved with a traditional CCD, readout noise is a serious hindrance for high frame rate imaging of faint targets.
One possible solution is the electron multiplying CCD (EMCCD), also known under the trade name L3CCD.

The CCD used in the SONG project (\url{astro.phys.au.dk/SONG}) is an EMCCD implemented in an Andor iXion$^{EM}$+ 897 camera with $16 \times 16 \mu m$ pixels; it is an electron multiplying frame transfer CCD. Compared to a conventional CCD, the serial register in an EMCCD has been extended. In the extended part of the register, the voltage  used to shift the captured electrons from pixel to pixel is not in the normal 5 V range, but on the order of 40 V. Consequently the probability that an electron will knock another electron out of a bound state has been dramatically increased, in a process know as impact ionization. Such an event will effectively multiply the electron similarly to the process in an avalanche diode or a photomultiplier. The details of electron multiplication in this particular camera have been described in \cite{emccd}.

Due to the stochastic nature of the impact ionization events, the number of electrons in a cascade from one photo electron is not constant; that is, the gain of the electron multiplying register is essentially random, in a similar way to an avalanche diode { or the dynodes in a photomultiplier.} This leads to a number of complications, but allows EMCCDs to produce images at very high readout speeds, even at very low light levels, without being dominated by readout noise. This makes them ideally suited for high frame rate applications. Because the analogy between the photomultiplier and the EMCCD, much of the statistics developed for the photomultiplier can be reused, when dealing with an EMCCD. The possibility of doing high frame rate imaging, like lucky imaging has been explored in numerous other articles and theses; see for instance \cite{luckyimaging1,luckyimaging2}

While the groundbreaking improvements achieved in spatial resolution by use of high frame rate techniques is thoroughly described in the scientific literature, little work has been devoted to the aspect of photometric capability, in terms of accuracy and stability of EMCCDs and high frame rate imaging.

In the application of EMCCD cameras to follow-up observations of gravitational microlensing events in the search for exoplanets, both spatial resolution and photometric accuracy is of paramount importance for results. In the following we therefore present our first results from analysis of the photometric quality of sequences of shifted and added EMCCD images of a very crowded stellar field.

\section{Determination of Bias and Spurious Charge}

{ The bias of a CCD is usually assumed to be composed of a fixed bias pattern over the pixel coordinates and a bias DC level which is common to all pixels. The DC level is usually some function of time. With a conventional CCD camera, we may normally assume that over a set of bias frames, corrected for bias DC level drift, the ADU values for each individual pixel will be normally distributed around the bias. We can therefore apply the mean of a set of bias frames as a good estimate of the true bias. This is not the case with our EMCCD camera because of the EM cascade amplifier. Also in a conventional CCD the bias DC level will usually only be weakly variable, because the temperature of the cooled CCD and on-chip readout amplifier is under servo control and therefore stable. Due to the readout speed and comparable large current through the EM cascade stages, there is a significant on-chip heat dissipation. One may therefore expect to see an appreciable bias DC level drift in an EMCCD camera, once the readout is initiated.}

\subsection{Exponentially Distributed Output from the EM Amplifier}

\begin{figure}
\begin{center}
\includegraphics[width=\columnwidth]{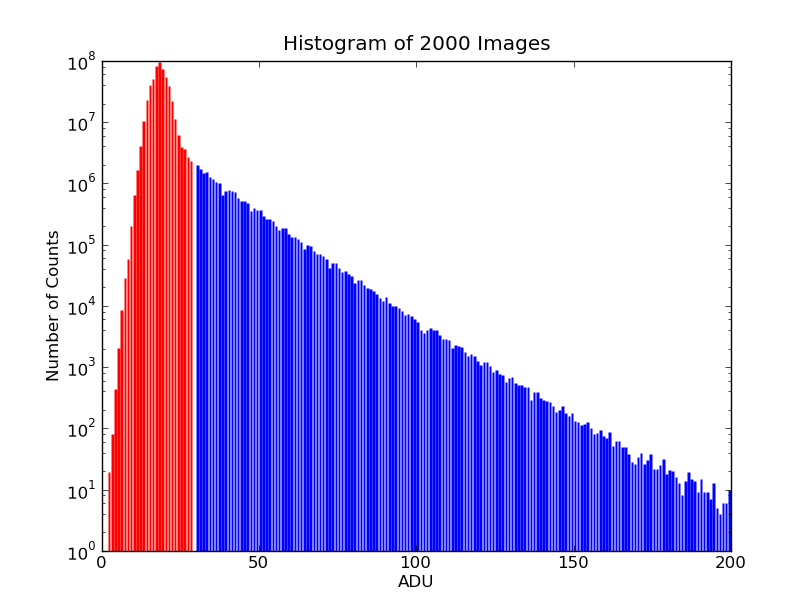} 
\end{center}
\caption{The histogram of 2000 dark images on a log scale. The exponential tail from the spurious charge can easily be seen. {For this particular example} the sum of the number of counts of the blue columns is approximately $4\%$ of the sum of the red columns. { The scale length of the tail is related to the electron multiplication gain, in accordance with Eq. \ref{expodist}}}
\label{hist} 
\end{figure}

Fig. \ref{hist} shows the histogram of 2000 dark frames from our high speed EMCCD camera, corrected for bias as described below. Here we see a classical Gaussian peak, corresponding to the well-known classical readout noise. Furthermore we see an exponentially decreasing tail to the right. { This tail arises from the EM cascade stages. Spurious electrons will arise randomly in the image array, even without exposure to light, as an effect of the parallel shifts and serial shifts in the EM register. They do also occur in a conventional CCD, but here they are undetectable because of the readout noise. But in an EMCCD spurious electrons will be cascade-amplified in the EM stage, giving rise to the exponential tail. Because of this peculiar distribution, the traditional mean value is not useful for determining the bias, which is the mode of the distribution in Fig. \ref{hist}. Because the distribution is asymmetric the mean is not an accurate estimator of the mode. Also, bias is not an integer, which implies that we cannot simply select the most common value in a histogram as the bias. But by appropriate truncation we can make the distribution approximately symmetric, hence we can use a truncated mean as an estimator of the mode.}

The truncated mean of a set of numbers is defined as the ordinary mean of the set where some percentage of the highest and/or lowest values has been discarded. Approximately $\approx 5-10\%$, depending on the gain and timing settings, of the pixels are affected by spurious charge and the affected pixels always have higher counts than the true bias. We will therefore for our purposes define the truncated mean as the mean of a set in which the $5\%$ highest values have been discarded, as this will exclude all amplified spurious electrons, assuming the rate of spurious electrons is constant.
{ Furthermore, a significant benefit of this method is that it is computationally fast.}
For a detailed description of EMCCD output, see \cite{emccd}.

\subsection{Bias DC Level Drift}

\begin{figure}
\begin{center}
\includegraphics[width=\columnwidth]{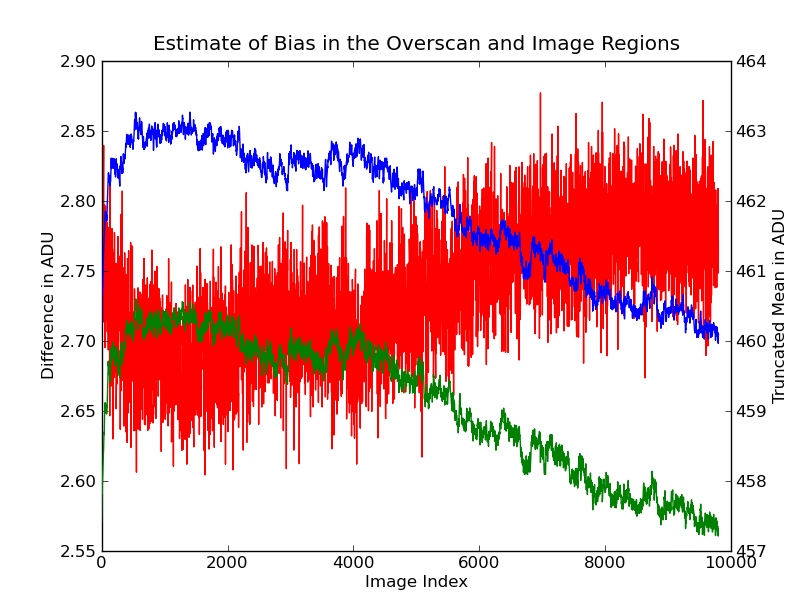}
\end{center}
\caption{Plot of the truncated mean of a time series of 10000 consecutive images. Blue is the truncated mean of the image area, and green is the truncated mean of the overscan region. Red is the difference between the two curves. An absolute variation in bias of about $\pm 3 ADU$ is seen over time (image index) for both the image area and the overscan region, whereas the variance of difference between them (red curve) is only about $\pm 0.05ADU$.}
\label{biasdrift}
\end{figure}

The CCD used in this experiment consists of an image area and two overscan regions, so that the first 20 pixels and the last 6 pixels of the 538 pixels in each row on the CCD are overscan regions, where no light reaches the CCD. 
In figure \ref{biasdrift}, the truncated mean of the overscan and the image area in a time series of 10000 images is shown. { It can be seen that there is considerable bias DC level drift over the course of the series for which one ought to correct. It can also be seen that the truncated mean of the overscan and image area varies in almost perfect unison. The truncated mean is needed as the overscan regions also have cascade amplified spurious electrons. The correlation coefficient between the two curves is 0.9995. Altogether, the truncated mean of the overscan appears to be a good estimator for the bias DC level of the image area, albeit with a constant offset of approximately 2.75.}

\subsection{Fitting Fixed Pattern Bias and Spurious Charge Rates}

{ In a CCD camera there is usually some fixed bias pattern in the image; this is also the case for this camera. Traditionally one takes the mean over several bias frames to obtain a good estimate of this bias pattern. Because of the bias drift and the spurious charges in the particular case, things have to be done a little differently. In long series, one would also like to be able to correct for the systematic background from the spurious charges. 

We therefore assume that the instantaneous bias $b_{inst}$ in the images can be written as
\begin{equation}
b_{inst}(x,y,t) = b(x,y) + b_{o}(t)
\end{equation}
That is, the instantaneous bias is composed of a fixed bias pattern $b$ that depends on the image coordinates, and a bias DC level $b_{o}$ that depends on time.}
For a set of bias frames $\{b_{inst}(x,y,t_i)\}$, a new set of bias frames is generated, where the DC bias level is normalised to zero.
\begin{equation}
\{b_c(x,y,t_i)\} = \{b_{inst}(x,y,t_i) - \underset{(x,y)}{\mu[5\%]}(b_{inst}(x,y,t_i))\}
\end{equation}
where $\underset{(x,y)}{\mu[5\%]}$ is the 5\% truncated mean from above over the pixel coordinates $x$ and $y$. 
To estimate $b$ it will assumed that there is at most one spurious electron per pixel per readout in a series of bias frames. It is also assume that the size of the electron cascade arising from one electron through the EM multiplier $X$ is given by an exponential distribution:
\begin{equation}
P(X=x) = {\gamma} e^{-{\gamma}x} H(x)
\label{expodist}
\end{equation}
where $H$ is a Heaviside function and $\gamma$ is the EM amplification.
In the case that no electron entered the EM multiplier we assume a constant bias reading; that is, the probability distribution of a bias $B$ is given by
\begin{equation}
P(B=x) = \delta(x)
\end{equation}
An EMCCD still has conventional additive Gaussian readout noise, but this noise is added after the EM multiplication. We will define $R$ as a random variable representing Gaussian readout noise around the bias value $b$, the numerical value of $b$ being a property of the readout electronics, that is
\begin{equation}
P(R=x) = \mathcal{N}(x-b,\sigma)
\end{equation}
where $\mathcal{N}$ is the normal distribution probability density function (PDF):
\begin{equation}
\mathcal{N}(x,\sigma) = \frac{1}{\sqrt{2 \pi \sigma^{2} } }e^{-\frac{(x)^{2}}{2\sigma^{2}}}
\end{equation}

Having corrected all the frames for a bias DC level we will therefore assume the following total outcome $Z$ of a "bias" reading
\begin{equation}
Z = \left\{ \begin{array}{rl}
 B &\mbox{ with $p$} \\
  X &\mbox{ with $1-p$} \\
       \end{array} \right\} + R
\end{equation}
Since $B$ and $X$ are mutually exclusive, and the PDF of a sum of random variables is given by the convolution of the constituting probability distributions, we can write
\begin{align}
\label{zdist}
P(Z=n) =  & \int \left( p\delta(\xi) + {(1-p)}{\gamma} e^{-\gamma \xi} H(\xi) \right) \\ & \mathcal{N}(n-b-\xi,\sigma) d\xi \nonumber
\end{align}
where $\mathcal{N}$ is the normal distribution PDF, and $1-p$ is the probability of a spurious charge. This equation is a mixture distribution of a zero output representing the event of no spurious electron and an exponentially distributed output representing the event of a spurious electron. All the parameters are to be considered functions of the pixel coordinates $x$ and $y$.

We will ignore the width of the normal distribution when convolving with the exponential distribution, because the breadth of the exponential distribution is much larger than the normal distribution if the EM gain is high. This allows us to write Eq. \ref{zdist} as
\begin{equation}
P(Z=n) \approx p\mathcal{N}(n-b,\sigma) + (1-p) \gamma e^{-\gamma (n - b) } H(n-b)
\label{photon}
\end{equation}
This derivation can be extended into a compelling method for counting photons in data from EMCCD; see \cite{emccd}.

The standard method for fitting a mixture distributions is the Expectation Maximization (EM) algorithm \citep{Dempster77}, not to be confused with the abbreviation for Electron Multiplying. This algorithm is a two step iterative algorithm, consisting of an expectation step and a maximization step. In the expectation step the probability of each data point belonging to each of the two component distributions is estimated according to
\begin{equation}
\lambda_i = \frac{\hat{p} \mathcal{N}(n_{i}-\hat{b},\hat{\sigma})}{(1-\hat{p}) \hat{\gamma} e^{\hat{\gamma}(n_{i}-\hat{b})} + \hat{p}\mathcal{N}(n_i-\hat{b},\hat{\sigma})}
\end{equation}
where $\lambda_i$ is the posterior probability that the i'th reading is bias. Hence $(1-\lambda)$ is the probability that the reading is due to a amplified spurious electron. $E$ is the exponential distribution.
In the subsequent maximization step, the weighted maximum likelihood estimate of all the parameters is calculated. 
\begin{equation}
\hat{p}_{k+1} = \frac{1}{m} \sum_{i} \lambda_{i}
\end{equation}
\begin{equation}
\hat{b}_{k+1} = \frac{\sum_{i} \lambda_{i}(n_{i} - \hat{b}_{k})}{\sum_{i} \lambda_{i}}
\end{equation}
\begin{equation}
\hat{\sigma}_{k+1} = \sqrt{\frac{\sum_{i} \lambda_{i}(n_{i} - \hat{b}_{k})^{2} }{\sum_{i} \lambda_{i}}}
\end{equation}
\begin{equation}
\hat{\gamma}_{k+1} =  \frac{\sum_{i} (1-\lambda_{i}) }{\sum_{i} (1-\lambda_{i}) (n_{i} - \hat{b}_{k})}
\end{equation}
where $\hat{b}_{k}$ is the estimated bias from the previous iteration.

\begin{figure}
 \begin{center}
 \subfigure[Bias pattern]{\includegraphics[width=0.45 \columnwidth]{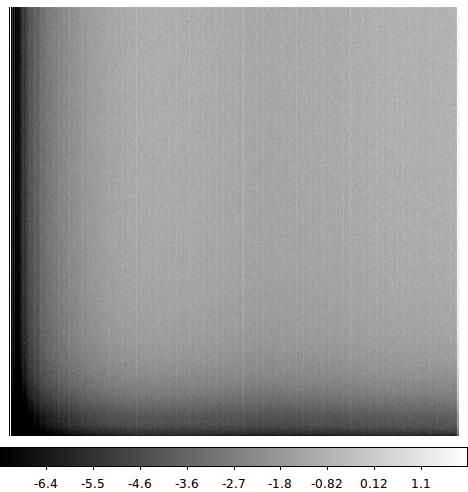}}
 \subfigure[Spurious charge]{\includegraphics[width=0.45 \columnwidth]{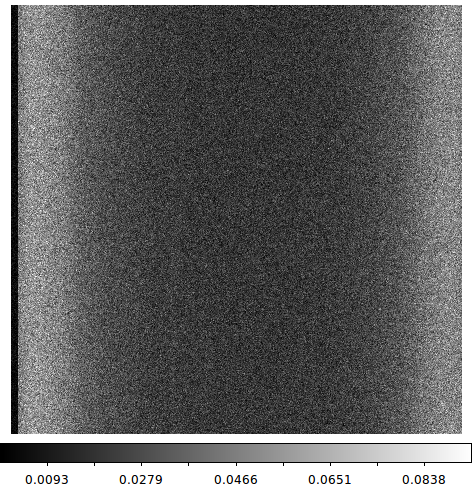}}
 \subfigure[EM gain]{\includegraphics[width=0.45 \columnwidth]{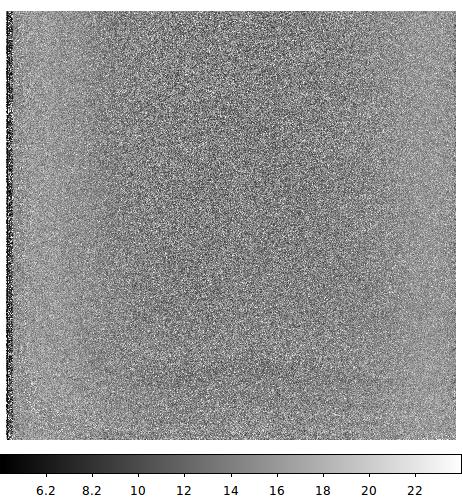}}
 \subfigure[Read out noise]{\includegraphics[width=0.45 \columnwidth]{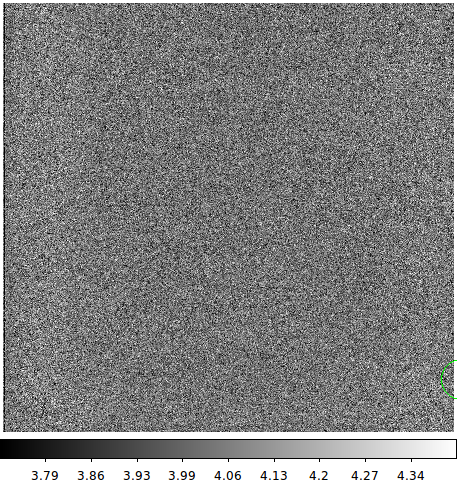}}
 \end{center}
 \caption{The results of running the proposed EM algorithm on a stack of 500 DC level corrected bias frames. There is a clear structure in the fixed bias, the structure has an 8 pixel modulation, presumably from the read out amplifier. The spurious charge probability map shows a valley-like structure. This structure is expected as the clock pulses for vertical shift will be smoothed out passing through the chip. Also the spurious charge probability in the overscan region to the left is very low as these pixels are virtual. The map of the EM gain and the read out noise show very little structure which is to be expected, because there is only one readout amplifier. There is a slight structure in the gain map. The structure is due to less variance in the gain determination at the edges, as the rate of spurious electrons, which carry information about the gain, is higher at the edges.}
\label{bias}
\end{figure}

By running this iteration to convergence for all pixel coordinates x and y in a stack of DC level corrected bias frames, maps of the spurious charge probability, EM gain, readout noise and bias can be obtained, as illustrated in Fig. \ref{bias}. 

It can be seen that there is a clear structure in the spurious charge probability distribution. This pattern is a logical consequence of the way this CCD is clocked. As the clocking pulse travels over the array of pixels, the train of pixels functions as a low pass filter, smoothing out the edge of the pulse this leads to a lower rate of voltage change (current) over pixels closer to the center, hence the lower rates of spurious electrons. This pattern is not a major concern for relative photometry of point sources, but for long exposures of extended sources with this type of camera, the effect will have to be corrected for.  
 
For a given raw science image $c(x,y,t)$, composed of the real image $c_b$ and the bias $b$ we can then calculate the DC level as
\begin{equation}
b_{o}(t) = \underset{(x_{o},y_{o})}{\mu[5\%]}(c(x,y,t) - b(x,y))
\end{equation}
where $x_o$ and $y_o$ is the pixels coordinates of the over scan region, because the DC level $b_{o}(t)$ is only a function of time, not $x$ and $y$. { Note that the offset between the overscan region and the image area has been absorbed into $b$}.
Finally we can correct the image as
\begin{equation}
c_b(x,y) = c(x,y,t) - [ b(x,y) + b_{o}(t) ] - (1/\gamma) s(x,y)
\label{final_corr}
\end{equation}
where $s = 1-p$ is the probability of a spurious charge and $1/\gamma$ is the average EM gain. Strictly the correction term $s$ will only correct for background from parallel clock induced charge and not clock induced charge in the serial registers, but as the serial register is common to all pixels this background contribution is not expected to be a function of the pixel coordinates and therefore not important to correct for. 

\begin{figure}
\begin{center}
\includegraphics[width=\columnwidth]{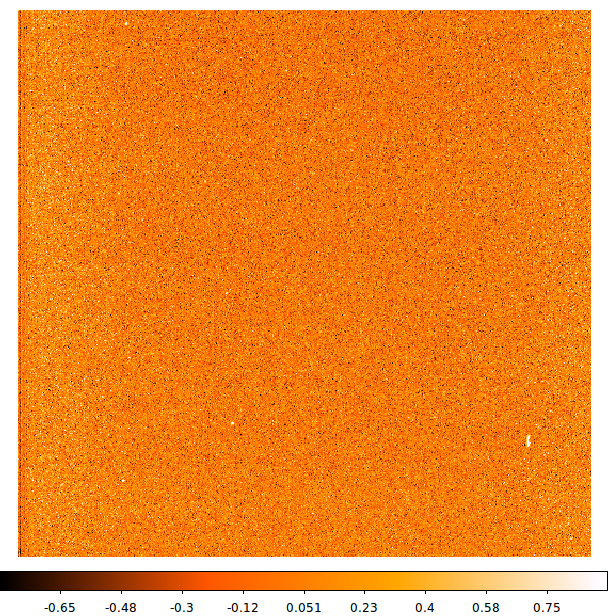}
\end{center}
\caption{Average of 10,000 reduced images, the image is reasonably flat, the variance of the image area is approximately 0.06 ADU}
\label{average}
\end{figure}

{ As a check on the consistency of the procedure outlined above, the average of 10,000 bias frames is presented in Fig. \ref{average}. As it can be seen from the spurious charge map, the rate of spurious charges per pixel per readout changes systematically over the image from about 2\% to about 8\%. Furthermore, with the applied settings on the camera we found the average EM gain ($1/\gamma$) to be $20.9ADU/e^{-}$. Over many frames this will average to a systematic background with a peak to valley range of approximately $(8\% - 2\%)(1/\gamma) = 1.3$ ADU per frame. Reducing the frames as outlined above, by subtracting the spurious charge rate map suitably scaled as in Eq. \ref{final_corr}, should remove this background. As there is no structure in the y direction of the spurious charge rate map, this map was smoothed by averaging in the y direction. 
The average of these images is shown in figure \ref{average}: the procedure seems consistent, the image is flat, and the mean value is close to zero. The variance of the image area is approximately 0.06 ADU. Over 10,000 frames, the expected variance given Eq. \ref{bias_var} would be approximately 0.004 ADU. This suggests that the noise is dominated by systematics. In fact, the noise in sums of more than about 1000 empty frames seems to be dominated by systematic noise.}

Flat fielding with EMCCDs is foreseen to be analogous to conventional CCDs; it corrects for differences in sensitivity between pixels and dust in the optical train. These effects affect an EMCCD in the same way as a conventional CCD.

\subsection{Noise Scaling}

{ The standard noise model for conventional CCD does not apply to EMCCDs. In most cases the readout noise can be ignored, but the cascade amplifier effectively doubles the photon noise, which is equivalent to cutting the quantum efficiency in half, if photon counting is not performed. 

Specifically, if the cascade amplifier is viewed as a linear amplifier, the signal is the mean value of a mixture distribution like Eq. \ref{photon} , which is simply the weighted average. I.e., the mean value of Eq. \ref{photon} is
\begin{equation}
E(Z) = b + \frac{1-p}{\gamma}
\end{equation}
In general the $j$'th central moment of a mixture distribution is given by \citep[pg. 11]{mixture}
\begin{equation}
E((Z-\mu)^j) = \sum_{i}\sum_{k=0}^{j} \left(  \begin{array}{c} j \\ k \end{array}   \right) (\mu_i-\mu)^{j-k} p_i E((X_i-\mu_i)^{k})
\label{mixture}
\end{equation}  
where $\mu$ is the mean value of the mixture distribution and $\mu_i$ is the mean values of the component distributions. 
Knowing the variance and mean of the exponential and normal distributions, one can then calculate the variance or second central moment of Eq. \ref{photon}. Assuming a bias of $b=0$ the expression simplifies to 
\begin{equation}
E((Z-E(Z))^2) = p\sigma^2 + \frac{1-p^2}{\gamma^{2}}
\label{bias_var}
\end{equation} 
In deriving Eq. \ref{photon} we have explicitly assumed that at most one electron gets cascade amplified. In the more general case of higher fluxes one cannot ignore coincident photoelectrons. According to \cite{emccd}, the distribution of the output can be generalized to 
\begin{equation}
P(Z=n) \approx p_0\mathcal{N}(n,\sigma) + \sum_{i=1}^{\infty} p_i\frac{\gamma^{i} n^{i-1}e^{-\gamma n  }}{(i-1)!}
\end{equation}
assuming $b=0$. The rightmost term is the PDF of the Erlang distribution and the $p_i$s are given by the Poisson PMF
\begin{equation}
p_i = \frac{\beta^{i}e^{-\beta}}{i!}
\end{equation}
One sees that the approximation in Eq. \ref{photon} is adequate when $\beta$ is small so that $p_i \approx 0$ for $i>1$, because the Erlang distribution for $i=1$ is the exponential distribution. Formally the factor $1-p$ in Eq. \ref{photon} refers to the spurious charge rate, but spurious charges will be Poisson distributed, hence they can be viewed as constant addition to the rate parameter for any photon flux. Thus, in this limit, the noise is given by Eq. \ref{bias_var}. The S/N ratio in this limit can be approximated as 
\begin{equation}
S/N = \frac{\beta+\alpha}{\sqrt{\sigma^{2}\gamma^{2}+2(\beta+\alpha)}}
\end{equation}
where $\alpha$ is the rate of the spurious charges.

If $\beta$ is large then $p_0 \approx 0$ and one obtains a different scaling for the noise. Specifically because the mean variance of the Erlang distribution for a given $i$ is $i/\gamma$ and $i/\gamma^2$, one finds that 
\begin{equation}
E(Z) = \sum_{i=1}^{\infty} p_i \frac{i}{\gamma} = \frac{\beta}{\gamma}
\end{equation}
because the mean value of the Poisson distribution is exactly $\beta$.
Further one can calculate the variance according to Eq. \ref{mixture}
\begin{align}
E((Z-E(Z))^2) = & \frac{1}{\gamma^2} \left( \sum_{i=1}^{\infty} p_i (i - \beta)^2 + \sum_{i=1}^{\infty}p_i i \right) \\ = & \frac{1}{\gamma^2} (\beta + \beta) \nonumber =  \frac{2\beta}{\gamma^2}
\end{align}
the fist term in the sum evaluates to $\beta$ because the variance of the Poisson distribution is $\beta$, and this term is by definition the variance. Calculating the S/N ratio in this limit we find that
\begin{equation} 
S/N = \frac{\beta / \gamma}{\sqrt{2\beta /\gamma^{2}}} = \frac{\beta}{\sqrt{2\beta}}.
\end{equation}}

\section{The Image Registration Algorithm}

In an astronomical image taken through a Kolmogorov atmosphere the dominant  seeing aberrations in terms of Zernike polynomials are piston, tip and tilt. For imaging purposes the zero'th term piston (i.e. an overall phase delay) is of no importance. We will therefore try to design the algorithm to correct for tip and tilt as fast, efficiently and accurately as possible.

\subsection{Correcting tip and tilt}

To utilize the signal in the whole image, we propose using the Fourier cross correlation theorem for correcting tip and tilt, which translates to an overall solid body translation of the image. { This method is different from the more common method of registering the frames based on one or more centroids \citep{luckyimaging1,luckyimaging2}.}

Given a set of bias- and flat field- corrected images $\{I_{i}(x,y)\}$, as described earlier, we will generate a reference image $R$ by taking the average.
\begin{equation}
R(x,y) = \frac{\sum_{i=1}^{N}I_{i}(x,y)}{N}
\end{equation}
The mean image represents the mean position of the image, and we will then shift all the individual images to this position and co-add them.
A method for finding the shift between two images using fast Fourier transforms (FFT) has been described by \cite{Araiza08}. It can be shown that the shift between two images where
\begin{equation}
I_i (x,y) = R (x + \Delta x_i, y+\Delta y_i)
\end{equation}
can be found using the following expression for the cross correlation between the images:
\begin{equation}
P_{i}(x,y) = \vert FFT^{-1} [ FFT(R) \cdot \overline{FFT(I_i )} ] \vert
\end{equation}
This expression will have one global maximum at $(\Delta x_i, \Delta y_i)$. The appropriate shift can be found by looking for the position of the global maximum in $P_i$.

Using the Fourier cross correlation theorem implicitly assumes that the images are circularly shifted. This is obviously not the case and some sort of apodisation of the images is called for to avoid edge effects. Furthermore in the field of adaptive optics the size of the typical isoplanaric patch is on the order of 10\arcsec. One would therefore expect the size of of a ''lucky'' patch to be similar. Surprisingly experience shows that the ''lucky'' patch size, where we do not see differential image motion, is significantly larger. The patch size is on the order of 35\arcsec, which is slightly smaller than the field of view in our camera. We can therefore solve both the apodisation problem and the problem with differential image motion by multiplying by an appropriate apodisating Hamming window.  

\subsection{Instantaneous Image Quality}

The most widely used measure of instantaneous image quality in lucky imaging is simply the maximum value of the pixel values in the frame \citep{brightestpixel}. Over long time scales this measure suffers undue interference with fluctuations in atmospheric extinction and scintillation, simply because the maximum value scales with a multiplicative constant:
\begin{equation}
\max(aI_i + b) = a \max(I_i) + b
\end{equation}
We have therefore adopted another measure for instantaneous image quality based on $P_i$. Because the FFT is linear we have that
\begin{align}
P_{i}'(x,y) & =  \vert FFT^{-1} [ FFT(R) \cdot \overline{FFT(a I_i + b )} ] \vert \\ & = aP_{i}+b\vert R \vert
\end{align}
If we therefore adopt the maximum of $P_i$ scaled with its surroundings as a quality measure
\begin{equation}
q_i = \frac{P_i(x_{max},y_{max}) }{\sum_{\substack{\vert(x,y)\vert<r \\ (x,y)\neq (x_{max},y_{max})}} P_i(x,y)}
\end{equation}
any scaling factor $a$ will cancel out. This measure is sensitive to any offset $b$. It is therefore important to accurately calibrate out any offsets. This is also true for the more common maximum value measure.

This expression is a proxy for the breadth of the maximum correlation peak. Intuitively if a frame has high image quality it will fit (correlate) well in very narrow range of offsets, and if a frame has low image quality it will correlate less well over a broader range of offsets.

In \cite{staley} it is proposed that a LI PSF is a convex linear combination of a diffraction limited core and a diffuse halo in the form of a Moffat function akin to the conventional seeing disk. Consequently it would be rational to measure the instantaneous image quality as the relative weighting between these two components as a form of pseudo Strehl ratio. 

We have therefore tested whether the instantaneous image quality measure proposed here is a reasonable proxy for the relative weighting of the two PSF components. To this end we generated a 100 pixel 1D reference image with two PSF consisting of a Gaussian core with a FWHM of 2 pixels and a Moffat halo with a FWHM of 10 pixels. The  weight of the two components was 50\%:50\%. We then calculated the proposed image quality with respect to images where the weighting was varied from 100\%:0\% to 0\%:100\%, and found that the proposed quality measure was a monotonous approximately linear function of the relative weighting. We also found that if the reference was sharper, i.e. more weighted towards the Gaussian core, the relation between the quality measure and the weight ratio was steeper. This indicates that the image quality measure is more discriminatory when given a sharper reference image.

\section{The Implementation}

\subsection{The Camera and Optical Setup}
The lucky imaging system has been implemented on the Danish 1.54m Telescope { at the ESO La Silla Observatory in Chile. The camera used for the implementation is an Andor Technology iXon+ model 897 EMCCD camera. This camera has an image area of 512x512 $16\mu m$ pixels, corresponding to a pixel scale of 0 \farcs09 on sky. The average seeing at La Silla is around 1\arcsec. 

This system is intended as a testbed for the lucky imaging system of the SONG telescope network \citep{song}, and the specifications are therefore identical to SONG.} The firmware of the camera was specially modified by Andor to also read out the overscan regions; 20 columns to the left and 6 columns to the right of the image area. { The camera is equipped with two readout amplifiers, one conventional and one electron multiplying. For the lucky imaging experiments the camera is read out using the electron multiplying readout amplifier at a rate of 10MHz. In this mode the gain of the readout amplifier is $25.8\tfrac{e_{EM}^{-}}{ADU}$, with a readout noise of $65.8e_{EM}^{-}$, but this conversion takes place after electron multiplication stages, which, in this experiment, amplifies one photoelectron into approximately 300 electrons on average. Thus the formal readout noise is subelectron, on the order of $65.8e^{-}_{EM}/300\tfrac{e^{-}_{EM}}{e^{-}_{phot}}=0.2e^{-}_{phot}$. The notation with $e_{phot}^{-}$ and $e_{EM}^{-}$ is to highlight the difference between electrons before and after the cascade amplifier, respectively. 

One has to keep in mind that the output distribution of the electron multiplier is exponential, not normal, as this leads to the extra photon noise described previously. This particular setup translates one photon electron to $300\tfrac{e^{-}_{EM}}{e^{-}_{phot}}/25.8\tfrac{e^{-}_{EM}}{ADU} = 11.6\tfrac{ADU}{e^{-}_{phot}}$. Most photometry packages take a $gain$ input parameter, and assume that the noise is $\sqrt{signal/gain}$. One can take the extra photon noise into account by defining a new fictitious gain, $emgain = gain/2$. Inputting this gain will make the noise assumption read $\sqrt{signal/emgain} = \sqrt{2*signal/gain}$.} 

To communicate with the camera we used a dual core 3GHz PC with 2GB of memory and an iSCSI 600GB RAID0 array for intermediate data storage. 
On the PC we ran Ubuntu 9.10 as Andor delivers a Linux driver in the form of a .so shared library.

\subsection{Software implementation}
The software for handling the camera and reducing the data was written using Python, representing the images as NumPy arrays. 
The camera was controlled from Python using the andor.py project (\url{http://code.google.com/p/pyandor/}). This project wraps the .so driver via the Python extension ctypes, which makes it possible to control the camera and load the images directly from the camera as NumPy arrays. 
The driver implements a spool function that should be able to spool series of images as three dimensional FITS files directly to a disk, but this function was found to be non-functional when called from Python, for unknown reasons. Also the way a FITS cube is represented in a FITS file as a continuous binary blob without any indexing hampers performance and caching when manipulating such files.
 
Instead, a spool function was implemented using the PyTables \citep{pytables} project, thereby enabling the spooling of data directly to disk in the form of HDF5 files at a satisfactory rate.
The FFT for the lucky imaging reduction was done using functions from the FFTW3 library \citep{FFTW05}, linked into Python via the PyFFTW project, to get a satisfactory processing speed. For our purposes the FFTW3 FFT implementation proved to be some 20 times faster than the stock FFT from NumPy.

To be able to handle the data reduction and the data acquisition simultaneously, the software was designed to be multi threaded using the Python extension multiprocessing. The software runs a thread for handling the graphical user interface, a thread for controlling the camera and acquiring data, and threads for reducing data.

When the camera handling thread gets an order to acquire a lucky image, it creates a HDF5 \citep{hdf5} file, saves the current bias frame, flat frame, and the data about the setup into the file. It then spools the image data into the file. When done it starts an independent reductions thread with the name of the created HDF5 file as a parameter. The HDF5 file contains all the information the reduction process needs to create a reduced image. 

The output from the reduction is a FITS file with 10 images in it. The header of the file contains standard information about the reduction, and the camera setup. The first image in the FITS file is the shifted sum of the original images in the image quality ranking $q$ brackets. In this way the user can later choose the best ratio between image quality and signal to noise. The reduction software only shifts the images to an integer number of pixels to avoid interpolation.

\section{Lucky Imaging Photometry}

There is still much confusion about what the term lucky imaging implies. The original definition of
lucky imaging \citep{Fried78} is when, out of a stack of high frame rate images, only a small fraction of
images that are diffraction-limited, or near-diffraction-limited are kept. Depending in seeing conditions, wavelength and telescope aperture,
only of the order of 1\% of the images are kept.
These images are then shifted to correct for the most dominant atmospheric
aberration modes, tip and tilt, as these modes impart a solid-body shift that
can be corrected trivially. Finally the images are added to improve the signal.

This method will produce near-diffraction-limited images on 1-2m class
telescopes, albeit at the
cost of a high photon waste, which is in principle bad for time-resolved photometry. It is important to note
that the shifting and adding can be done for any image, regardless of the instantaneous image
quality. Simply shifting and adding all images in a high frame rate stack will usually improve the
seeing by a factor of approximately two, without any loss of photons. This method is known as shift-and-add.

\subsection{Strategies for photometry in crowded fields}

The problem of extracting photometry from an image is an archetypical inverse problem in the sense that it is easy to construct the image given the positions of the sources and the PSFs. But it is hard to extract the position of the sources and the PSFs given an image, especially in a crowded field. However, given the positions of all the sources in the image, the condition number of the inverse problem drops dramatically. One could therefore imagine a procedure for extracting time-resolved photometry where the positions of all the point sources in the image are extracted from a lucky image, but the time series of images is constructed from frames that are only shift-and-added, thus preserving all the photons with an improved resolution.

Another most interesting approach is differential image analysis (DIA), in which a high resolution reference image is blurred to the seeing and shift of images in the time series, and then subtracted. This method utilises information about the source distribution from the high resolution image. A lucky image constructed from a time series of observations seems ideally suited as a reference image in this respect. In particular, DIAs utilizing numerical kernels, such as DanDIA \citep{dandia}, seem well-suited because a { lucky imaging PSF in general seems to be peculiar and variable with a near diffraction limited core and an extended halo, comparable to the conventional seeing limit \citep{peculiar}.} In the vein of not wasting photons and cleverly including the high resolution information, another interesting technique is online deconvolution \citep{onlinedeconv}. These approaches will be pursued in future work. 

\subsection{Photometric Stability}

To test the stability of the photometry obtained from an EMCCD, along with the outlined reduction procedure, a 1.5 hour sequence of the center of the of the globular cluster \object{$\omega$ Cen} at the coordinates $13^{h} 26\arcmin 47 \farcs5 , -47\degr 28\arcmin 41 \farcs 0$ J2000, was acquired at a rate of 10 images per second. The images were then reduced, registered and had their image quality determined as outlined above. The observations were started at an hour angle of $1^{h} 33^{m}$ and an airmass of 1.1. This field centered on $\omega$ Cen was chosen to simulate the crowded conditions of microlensing observations towards the galactic bulge, and because there are extensive observations from the Hubble Space Telescope for comparison. 

\begin{figure}
\centering
\includegraphics[width=\columnwidth]{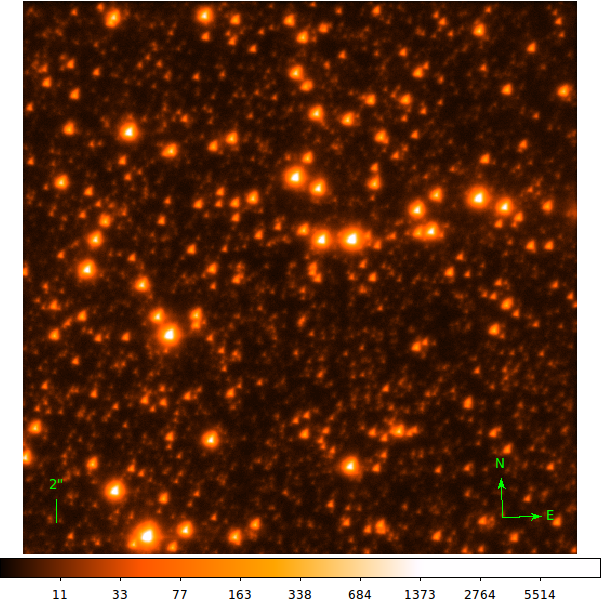}
\caption{Lucky Image constructed from the 1000 best seeing frames out of 50.000, determined by a simple brightest pixel criterion. The image has been reproduced on a log scale, to shown the extensive halos and the peculiar triangular core. The FWHM seeing is approximately 0\farcs4 and the pixels scale is 0\farcs09 per pixel.}
\label{omegacen0}
\end{figure}

\begin{figure}
\centering
\includegraphics[width=\columnwidth]{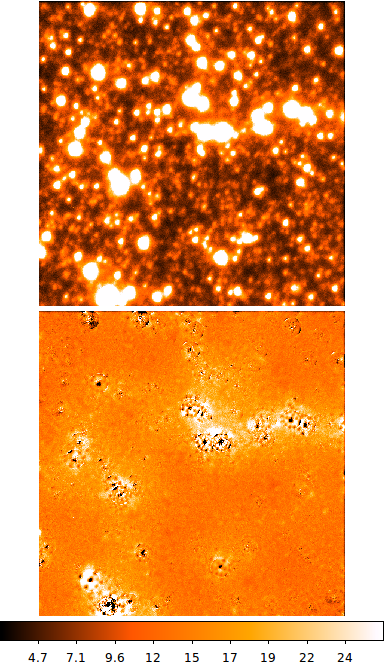}
\caption{The results of reducing a series of Omega Centauri with DaoPhotII. The upper image shows a typical single image composed of 500 shifted and added frames. The lower  image shows the residuals after subtraction with \textsc{DaoPhotII}.}
\label{substraction}
\end{figure}

This field is extremely crowded, and the improved resolution is a major virtue in finding the source positions in such a field. Hence the source positions were extracted with the standard PSF-fitting photometry package \textsc{DaoPhotII} \citep{Stetson87} from a lucky image composed of the 1\% sharpest images. This image has been reproduced in Fig. \ref{omegacen0}. It is evident that the resolution has been significantly improved compared with Fig. \ref{substraction}. 

Unfortunately the Danish telescope, commissioned in 1979, was never designed for imaging below the seeing limit.  Hence, the image is limited by triangular coma from the telescope at the $0 \farcs 3$ level, leading to peculiar triangular PSFs; see Fig. \ref{omegacen0}. It has proven to be difficult to extract an accurate PSF with \textsc{DaoPhotII} from this lucky image. LI PSFs generally have very extensive halos due to higher order atmospheric aberrations not corrected by lucky imaging. These very broad halos and the crowded field makes it very difficult not to pollute the wings of the \textsc{DaoPhotII} PSF with faints stars. { The diffraction limit for a 1.5m telescope in I is $0 \farcs11 $, but the intrinsic aberrations of the telescope imply that the images are not undersampled.

The inaccurate PSF determination leads to a less than optimal subtraction in \textsc{DaoPhotII}. But it should be noted that this image is only used to extract the positions of the stars, and it is still possible to extract accurate positions of even very faint stars from this image. }

To extract time series photometry, a sequence of 100 images were generated by shifting and adding 500 consecutive frames for an effective exposure time of 50s. The list of positions from the lucky image was then projected onto the 100 images with the \textsc{DaoPhotII} program \textsc{DAOMASTER}, and the full time-resolved photometry of 2523 stars was extracted simultaneously with the program \textsc{ALLFRAME} \citep{stetson94}.

In these 100 images, which have only been shift-and-added, the PSF is more conventional and the result of the subtraction is satisfactory, as seen in Fig. \ref{substraction}.

\begin{figure}
\centering
\includegraphics[width=\columnwidth]{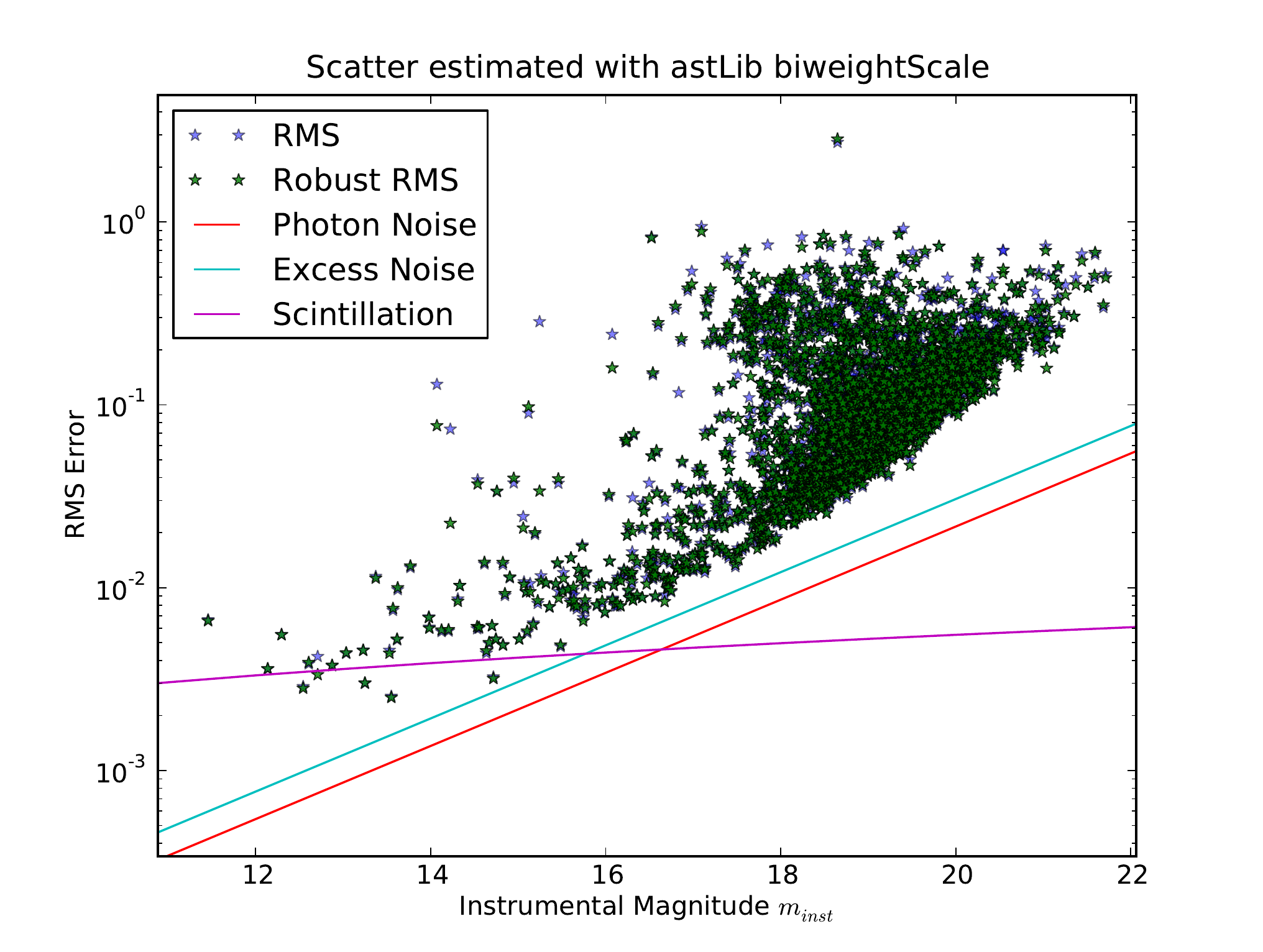}
\caption{The RMS scatter of the photometry of 2523 stars in the time series consisting of 100 images stacked from 500 0.1s exposures each. The photon noise limit and the excess noise limit has been plotted for comparison. Due to the stochastic amplification in the EM stages, an EMCCD will, when regarded as a conventional linear amplifier, effectively multiply the photon noise by a factor of 2. The robust RMS is estimated with the function biweightScale from the Python module \textsc{astLib}.}
\label{logScatter}
\end{figure}

To obtain relative photometry, the mean of an ensemble of 10 carefully selected stars was subtracted from the photometry at each time step. Further, for each of the 2523 light curves the RMS scatter was robustly estimated with the function biweightScale, from the \textsc{astLib} Python module. A plot of the relative error in this photometry is plotted in Fig. \ref{logScatter}. The magnitudes are the instrumental magnitudes reported by \textsc{DaoPhotII}. To simulate a SONG telescope a special longpass filter with a cut-on wavelength of 650nm (Thorlabs FEL0650) was used, hence the magnitudes are not directly translatable to the Johnson BVRI system. 

\begin{figure}
\centering
\includegraphics[width=\columnwidth]{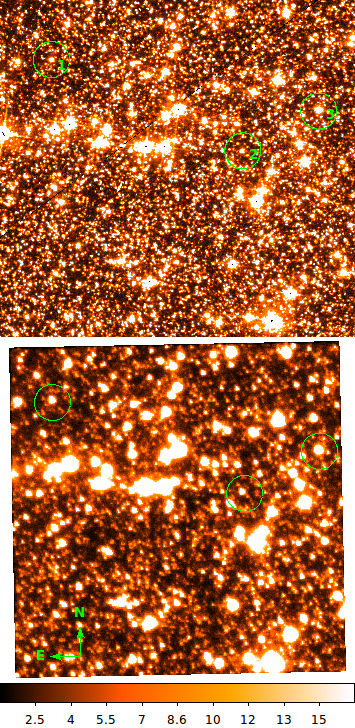}
\caption{Comparison of the image in Fig. \ref{omegacen0} with an image form Hubble Space Telescope. The three stars marked have been used to find the approximate offset in the photometric zero point between the STMAG system and the instrumental magnitudes.}
\label{hubbleim}
\end{figure}

We found that the field we have investigated was observed from the Hubble Space Telescope in 1997 with the WFPC2 camera in the F675W filter. From this we extracted the aperture photometry of three stars that were faint enough not to be saturated and reasonably isolated. The photometry extracted from the three stars in Fig. \ref{hubbleim} has been summarised in table \ref{phot3stars.tab}, converted to the STMAG system for the F675W filter, and approximated to Johnson R according to the instructions in the WFPC2 Photometry Cookbook.

\begin{table}
\label{phot3stars.tab}
\caption{Photometry of 3 selected stars}
\begin{center}
\begin{tabular}{c|c|c|c|c}
Index & STMAG & Johnson R &  $m_{\mathrm{inst}}$ & R - $m_{\mathrm{inst}}$\\ 
\hline 1 & 15.24 & 15.94 & 15.36 & 0.58\\ 
\hline 2 & 15.66 & 16.36 & 15.95 & 0.41\\ 
\hline 3 & 14.54 & 15.23 & 14.31 & 0.91\\
\hline avg & -   & -     & - &     0.63\\
\end{tabular}
\end{center}
\end{table}

For stars brighter than approximately $m_{\mathrm{inst}} = 16$, we see photometric scatter at a level consistent with scintillation. The scintillation level in Fig. \ref{logScatter} has been calculated according to Eq. 10 in \cite{scint}, assuming a telescope diameter of 154cm, a telescope altitude of 2340m and an airmass of 1.1.

For stars fainter than 16th magnitude but brighter than approximately 18th magnitude, the scatter seems to be bounded by photon noise to within 50\%. 

Finally for stars fainter than 18th magnitude we find a lower bound which rises more sharply than photon noise, plausibly because of the impact of crowding. 

\begin{figure}
\centering
\includegraphics[width=\columnwidth]{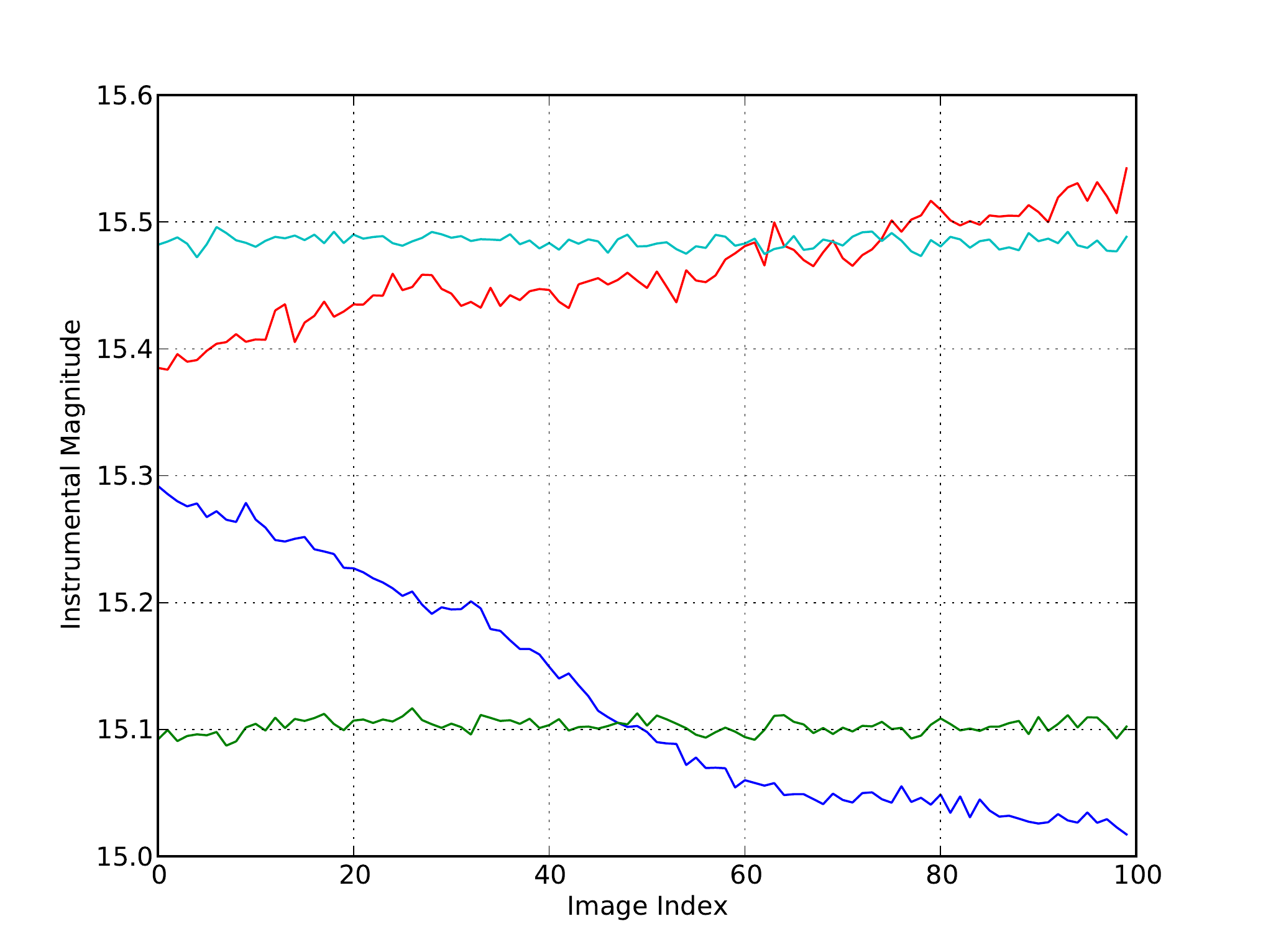}
\caption{For reference, two examples of constant light curves and two variable light curves extracted from the dataset have been plotted. Each image has a cumulated exposure time of 50s.}
\label{varstars}
\end{figure}

\section{Conclusions}
We have in this brief paper demonstrated that photometry in very crowded fields with high frame rate EMCCD data is indeed feasible.

We had to adjust conventional procedure for photometric data extraction to take account for the exponentially distributed EMCCD output and the bias variation, but worries about whether the stochastic nature of the EM amplification itself hinders accurate photometry over long time scales have been rebutted. In fact, the photometric scatter is close to the theoretical limits over most of the explored range of magnitudes.

For stars fainter than 19th magnitude we get a lower bound on the photometric scatter, which is worse than predicted from photon noise, plausibly due to crowding. We believe that this can be remedied to an extent by the virtue of being able to produce high resolution references from the data set and cleverly including this information via differential image analysis, DIA, or online deconvolution.

Finally we have demonstrated a very fast and robust implementation of the lucky imaging technique, based on cross correlation calculated in Fourier space. It is fast because it is based on FFTs and robust in the sense that no explicit reference stars have to be established, the algorithm is hands-off, and the whole image with all its information is utilised. This is most important in a robotic telescope network that is designed to observe microlensing events autonomously. Unfortunately it is difficult to find subpixel shifts with this method, but it is feasible and will be investigated in future work. { Because of intrinsic aberrations in the Danish telescope at present, the images are not undersampled, but they would have been if the telescope had been diffraction limited. In this case there will be information at high spatial frequencies that can be extracted by finding subpixel shifts and applying a dithering method. But even in the well-sampled case, super sampling, subpixel shifts, and dithering would produce more smooth PSFs which are potentially more easily handled with PSF fitting photometry software, hence potentially leading to more accurate photometry.}

\begin{acknowledgements}
The authors would like to acknowledge Anton Norup S\o rensen's thorough and diligent work designing, implementing and characterizing optics for the SONG project. The authors would also like to acknowledge valuable discussions with Preben N\o rregaard regarding the electrical engineering aspects of CCDs and EMCCDs. 
\end{acknowledgements}
\bibliographystyle{aa}
\bibliography{BiasDetermination}

\end{document}